\documentclass{achemso}
\setkeys{acs}{articletitle = true, maxauthors=0}
\SectionNumbersOn

\usepackage{chemformula} 
\usepackage[T1]{fontenc} 
\usepackage{multicol}
\usepackage{booktabs}
\usepackage{rotating}
\usepackage{amsmath}\allowdisplaybreaks[1]
\usepackage{braket}
\usepackage{xcolor}
\usepackage{mathrsfs}
\usepackage{simpler-wick}
\usepackage{cancel}
\usepackage{siunitx}

\def\hho{water}
\def\hhoo{(\textit{P})-hy\-dro\-gen per\-ox\-ide}
\def\smo{(\textit{S})-meth\-yl\-ox\-i\-rane}
\def\rcb{(\textit{R})-3-chlo\-ro-1-bu\-tene}
\def\rmo{(\textit{R})-4-meth\-yl-2-ox\-et\-a\-none}
\def\dma{(\textit{M})-1,3-di\-meth\-yl\-all\-ene}

\author{Brendan M. Shumberger}
\author{T. Daniel Crawford}
\email{crawdad@vt.edu}
\affiliation{Department of Chemistry, Virginia Tech, Blacksburg, Virginia 24061, U.S.A.}

\title{Analytic Computation of Vibrational Circular Dichroism Spectra Using Configuration Interaction Methods}

\abbreviations{AATs, CID, CISD}

\begin{document}

\begin{abstract}
In this work, we present the first derivation and implementation of analytic
gradient methods for the computation of the atomic axial tensors (AATs)
required for simulations of vibrational circular dichroism (VCD) spectra using
configuration interaction methods including double (CID) and single and double
(CISD) excitations.  Our new implementation includes the use of non-canonical
perturbed orbitals to improve the numerical stability of the gradients in the
presence of orbital near-degeneracies, as well as frozen-core capabilities.
We validated our analytic CID and CISD formulations against two new
finite-difference approaches.  Using this new implementation, we investigated
the significance of singly excited determinants and the role of CI-coefficient
optimization in VCD simulations by comparisons among Hartree-Fock (HF) theory,
second-order M{\o}ller-Plesset perturbation (MP2) theory, CID, and CISD
theories.  For our molecular test set including \hhoo, \smo, \rcb, \rmo, and
\dma\ we noted sign discrepancies between the HF and MP2 methods compared to
that of the new CID and CISD methods for four of the five molecules as well as
similar discrepancies between the CID and CISD methods for \dma.

\end{abstract}

\section{Introduction}

Vibrational circular dichroism (VCD), the differential absorption of circularly
polarized infrared radiation, is among the most widely used spectroscopic
techniques for determining the absolute stereochemical configurations of chiral
compounds.\cite{Stephens1985a, Nafie97, Freedman03, Barron04, Polavarapu07,
Crawford06:chirality, Crawford07:feature} However, unlike other chiroptical
spectroscopic methods such as optical rotation,\cite{Grimme01, Polavarapu02,
Ruud02, Norman04, Crawford08:optrot} electronic circular
dichroism,\cite{Carnell94, Hansen99, Berova00, Autschbach02:ECD, Diedrich03} and
Raman optical activity,\cite{Ruud02:ROA, Pecul05:ROA, Liegeois07, Crawford2011}
theoretical development of VCD simulations has, until recently, been limited
density-functional theory (DFT) methods.  

The principal challenge in implementing VCD within any quantum chemical
approach, whether Hartree-Fock (HF), DFT, or more advanced wave-function-based
methods, is that the magnetic dipole transition moment unphysically vanishes
within the Born-Oppenheimer approximation.  A generalized approach to treating
this problem was developed by Stephens in 1985 who proposed first-order
perturbations with respect to magnetic fields and nuclear displacements be
included in the ground-state wave function \cite{Stephens1985a}.  By relating
the first-order perturbations back to Taylor series expansions, the magnetic
dipole transition moment could be formulated as an overlap between ground-state
wave function derivatives with respect to nuclear coordinates and an external
magnetic field, known as the atomic axial tensor (AAT).  Shortly after
Stephens's breakthrough, the AAT was implemented using HF
theory\cite{Stephens1985b,Lowe1986,Amos1987}, multiconfigurational
self-consistent field (MCSCF) theory\cite{Bak1993}, and density functional
theory\cite{Cheeseman1996} bringing VCD to life in these methods.  However, for
some problems, higher accuracy was still desired and methods which mixed
calculations of the AAT, the atomic polar tensor (APT) (the second derivative of
the energy with respect to nuclear coordinates and an external electric field),
and vibrational Hessian were developed\cite{Stephens94:B3LYP}.

Due to the complicated formulation of the AAT, almost 30 years passed before
further implementation into wave-function-based methods that include dynamical
electron correlation, second-order M{\o}ller-Plesset perturbation theory (MP2)
and configuration interaction including double excitations (CID) using a
finite-difference approach\cite{Shumberger2024}.  Due to the impractical
computational scaling of numerical differentiation for wave functions involving
determinantal expansions [$O(N)^{11}$], Shumberger, Pearce, and Crawford
developed an analytic formulation based on derivatives of excited determinants
using a second-quantized approach, which they validated through the
finite-difference formulation reducing the computational scaling to $O(N)^5$
(given the appropriate intermediates) \cite{Shumberger2025}.

The purpose of the present work is to extend these methods for the first time to
the configuration interaction singles and doubles (CISD) method in order to
examine the significance of singly excited determinants on VCD rotatory
strengths.  In addition, we report additional improvements upon previous
implementations allowing the use of non-canonical perturbed
orbitals,\cite{Handy85:MP2Hessians} which provide greater numerical stability in
the case of near-degenerate orbitals, and frozen-core for greater computational
efficiency and for use with basis sets designed for the treatment of valence
correlation (e.g., the cc-pVXZ basis sets of Dunning and
co-workers.\cite{Dunning1989})

\section{Theory}

From Stephens's formulation of VCD, the magnetic dipole transition moment
includes the quantity $I_{\alpha\beta}^{\lambda}$, which is the electronic
contribution to the atomic axial tensor (AAT). This quantity is formulated as
the overlap between ground-state wave function derivatives where the bra-state
wave function is differentiated with respect to nuclear displacements,
$R_{\lambda\alpha}$, where $\alpha$ is a Cartesian coordinate of the $\lambda$-th
nucleus, and the ket-state wave function is differentiated with respect to the
$\beta$-th Cartesian direction of the magnetic field, $H_\beta$,
\begin{equation} \label{ElectronicAAT}
    I_{\alpha\beta}^{\lambda} = \left\langle
    \left( \frac{\partial \Psi_G(\vec{R})}{\partial R_{\lambda\alpha}} \right)_{R_{\lambda\alpha} = R_{\lambda\alpha}^0} \bigg|
    \left( \frac{\partial \Psi_G(\vec{R_0}, H_{\beta})}{\partial H_{\beta}} \right)_{H_\beta = 0} \right\rangle.
\end{equation}
By approximating the wave function, $\Psi_G$, as an intermediately normalized
CISD wave function,
\begin{equation} \label{CISDWaveFunction}
    \left| \Psi_G(\vec{R}, H_{\beta}) \right\rangle \approx \left| \Phi_0 \right\rangle + \hat{C}_1 \left| \Phi_0 \right\rangle + \hat{C}_2 \left| \Phi_0 \right\rangle,
\end{equation}
we can obtain the CISD AAT where the $\hat{C}_1$ and $\hat{C}_2$ operators are 
\begin{align}
    \hat{C}_1 &= \sum_{ia} c_i^a a_a^{\dagger} a_i \label{c1}
\end{align}
and
\begin{align}
    \hat{C}_2 &= \frac{1}{4} \sum_{ijab} c_{ij}^{ab} a_a^{\dagger} a_b^{\dagger} a_j a_i, \label{c2}
\end{align}
and $i, j, \ldots (a, b, \ldots)$ denote occupied (unoccupied) spin orbitals.
Insertion of Eq.~\eqref{CISDWaveFunction} into Eq.~\eqref{ElectronicAAT} and
evaluation of the operators leads to an expression for the AAT in terms of
CI coefficients, derivatives of CI coefficients, determinants, and derivatives
of determinants, i.e.
\begin{align} \label{CISDAATs1}
    [I_{\alpha\beta}^{\lambda}]_{int}
    & = \left\langle \frac{\partial \Phi_{0}}{\partial R_{\lambda\alpha}} \bigg| \frac{\partial \Phi_{0}}{\partial H_{\beta}} \right\rangle \nonumber \\
    & + \sum_{ia} \frac{\partial {c_{i}^{a}}^\dagger}{\partial R_{\lambda\alpha}} \left\langle \Phi_{i}^{a} \bigg| \frac{\partial \Phi_{0}}{\partial H_{\beta}} \right\rangle
      + \sum_{ia} {c_{i}^{a}}^\dagger \left\langle \frac{\partial \Phi_{i}^{a}}{\partial R_{\lambda\alpha}} \bigg| \frac{\partial \Phi_{0}}{\partial H_{\beta}} \right\rangle \nonumber \\
    & + \sum_{ia} \frac{\partial c_{i}^{a}}{\partial H_{\beta}} \left\langle \frac{\partial \Phi_{0}}{\partial R_{\lambda\alpha}} \bigg| \Phi_{i}^{a} \right\rangle
      + \sum_{ia} c_{i}^{a} \left\langle \frac{\partial \Phi_{0}}{\partial R_{\lambda\alpha}} \bigg| \frac{\partial \Phi_{i}^{a}}{\partial H_{\beta}} \right\rangle \nonumber \\
    & + \frac{1}{4} \sum_{ijab} \frac{\partial {c_{ij}^{ab}}^\dagger}{\partial R_{\lambda\alpha}}
        \left\langle \Phi_{ij}^{ab} \bigg| \frac{\partial \Phi_{0}}{\partial H_{\beta}} \right\rangle
      + \frac{1}{4} \sum_{ijab} {c_{ij}^{ab}}^\dagger
        \left\langle \frac{\partial \Phi_{ij}^{ab}}{\partial R_{\lambda\alpha}} \bigg| \frac{\partial \Phi_{0}}{\partial H_{\beta}} \right\rangle \nonumber \\
    & + \frac{1}{4} \sum_{ijab} \frac{\partial c_{ij}^{ab}}{\partial H_{\beta}}
        \left\langle \frac{\partial \Phi_{0}}{\partial R_{\lambda\alpha}} \bigg| \Phi_{ij}^{ab} \right\rangle
      + \frac{1}{4} \sum_{ijab} c_{ij}^{ab}
        \left\langle \frac{\partial \Phi_{0}}{\partial R_{\lambda\alpha}} \bigg| \frac{\partial \Phi_{ij}^{ab}}{\partial H_{\beta}} \right\rangle \nonumber \\
    & + \sum_{ia} \sum_{kc} \frac{\partial c{_{i}^{a}}^\dagger}{\partial R_{\lambda\alpha}} \frac{\partial c_{k}^{c}}{\partial H_{\beta}}
        \left\langle \Phi_{i}^{a} \bigg| \Phi_{k}^{c} \right\rangle
      + \sum_{ia} \sum_{kc} \frac{\partial {c_{i}^{a}}^\dagger}{\partial R_{\lambda\alpha}} c_{k}^{c}
        \left\langle \Phi_{i}^{a} \bigg| \frac{\partial \Phi_{k}^{c}}{\partial H_{\beta}} \right\rangle \nonumber \\
    & + \sum_{ia} \sum_{kc} {c_{i}^{a}}^\dagger \frac{\partial c_{k}^{c}}{\partial H_{\beta}}
        \left\langle \frac{\partial \Phi_{i}^{a}}{\partial R_{\lambda\alpha}} \bigg| \Phi_{k}^{c} \right\rangle
      + \sum_{ia} \sum_{kc} {c_{i}^{a}}^\dagger c_{k}^{c}
        \left\langle \frac{\partial \Phi_{i}^{a}}{\partial R_{\lambda\alpha}} \bigg| \frac{\partial \Phi_{k}^{c}}{\partial H_{\beta}} \right\rangle \nonumber \\
    & + \frac{1}{4} \sum_{ijab} \sum_{kc} \frac{\partial c{_{ij}^{ab}}^\dagger}{\partial R_{\lambda\alpha}} \frac{\partial c_{k}^{c}}{\partial H_{\beta}}
        \left\langle \Phi_{ij}^{ab} \bigg| \Phi_{k}^{c} \right\rangle
      + \frac{1}{4} \sum_{ijab} \sum_{kc} \frac{\partial {c_{ij}^{ab}}^\dagger}{\partial R_{\lambda\alpha}} c_{k}^{c}
        \left\langle \Phi_{ij}^{ab} \bigg| \frac{\partial \Phi_{k}^{c}}{\partial H_{\beta}} \right\rangle \nonumber \\
    & + \frac{1}{4} \sum_{ijab} \sum_{kc} {c_{ij}^{ab}}^\dagger \frac{\partial c_{k}^{c}}{\partial H_{\beta}}
        \left\langle \frac{\partial \Phi_{ij}^{ab}}{\partial R_{\lambda\alpha}} \bigg| \Phi_{k}^{c} \right\rangle
      + \frac{1}{4} \sum_{ijab} \sum_{kc} {c_{ij}^{ab}}^\dagger c_{k}^{c}
        \left\langle \frac{\partial \Phi_{ij}^{ab}}{\partial R_{\lambda\alpha}} \bigg| \frac{\partial \Phi_{k}^{c}}{\partial H_{\beta}} \right\rangle \nonumber \\
    & + \frac{1}{4} \sum_{ia} \sum_{klcd} \frac{\partial c{_{i}^{a}}^\dagger}{\partial R_{\lambda\alpha}} \frac{\partial c_{kl}^{cd}}{\partial H_{\beta}}
        \left\langle \Phi_{i}^{a} \bigg| \Phi_{kl}^{cd} \right\rangle
      + \frac{1}{4} \sum_{ia} \sum_{klcd} \frac{\partial {c_{i}^{a}}^\dagger}{\partial R_{\lambda\alpha}} c_{kl}^{cd}
        \left\langle \Phi_{i}^{a} \bigg| \frac{\partial \Phi_{kl}^{cd}}{\partial H_{\beta}} \right\rangle \nonumber \\
    & + \frac{1}{4} \sum_{ia} \sum_{klcd} {c_{i}^{a}}^\dagger \frac{\partial c_{kl}^{cd}}{\partial H_{\beta}}
        \left\langle \frac{\partial \Phi_{i}^{a}}{\partial R_{\lambda\alpha}} \bigg| \Phi_{kl}^{cd} \right\rangle
      + \frac{1}{4} \sum_{ia} \sum_{klcd} {c_{i}^{a}}^\dagger c_{kl}^{cd}
        \left\langle \frac{\partial \Phi_{i}^{a}}{\partial R_{\lambda\alpha}} \bigg| \frac{\partial \Phi_{kl}^{cd}}{\partial H_{\beta}} \right\rangle \nonumber \\
    & + \frac{1}{16} \sum_{ijab} \sum_{klcd} \frac{\partial c{_{ij}^{ab}}^\dagger}{\partial R_{\lambda\alpha}} \frac{\partial c_{kl}^{cd}}{\partial H_{\beta}}
        \left\langle \Phi_{ij}^{ab} \bigg| \Phi_{kl}^{cd} \right\rangle
      + \frac{1}{16} \sum_{ijab} \sum_{klcd} \frac{\partial {c_{ij}^{ab}}^\dagger}{\partial R_{\lambda\alpha}} c_{kl}^{cd}
        \left\langle \Phi_{ij}^{ab} \bigg| \frac{\partial \Phi_{kl}^{cd}}{\partial H_{\beta}} \right\rangle \nonumber \\
    & + \frac{1}{16} \sum_{ijab} \sum_{klcd} {c_{ij}^{ab}}^\dagger \frac{\partial c_{kl}^{cd}}{\partial H_{\beta}}
        \left\langle \frac{\partial \Phi_{ij}^{ab}}{\partial R_{\lambda\alpha}} \bigg| \Phi_{kl}^{cd} \right\rangle
      + \frac{1}{16} \sum_{ijab} \sum_{klcd} {c_{ij}^{ab}}^\dagger c_{kl}^{cd}
        \left\langle \frac{\partial \Phi_{ij}^{ab}}{\partial R_{\lambda\alpha}} \bigg| \frac{\partial \Phi_{kl}^{cd}}{\partial H_{\beta}} \right\rangle.
\end{align}
Using the orbital derivative approach developed for the MP2 AAT
\cite{Shumberger2025} and noting $\chi = R_{\lambda\alpha}$ or $H_\beta$, we
have that
\begin{align}
        \left| \frac{\partial \Phi_0}{\partial \chi} \right\rangle
    & = \sum_n U_{nn}^{\chi} \left| \Phi_0 \right\rangle
      + \sum_n \sum_f U_{fn}^{\chi} \left| \Phi_n^f \right\rangle
      + \sum_n a_{n_{\chi}}^{\dagger} a_n^{} \left| \Phi_0 \right\rangle. \label{ReferencedX}
\end{align}
In Eq.~\eqref{ReferencedX}, operators of the form $a_{p_{\chi}}^{\dagger}$
are second-quantized operators that create orbitals of the form
$\phi_p^{\chi}$, which denotes derivatives of the atomic-orbital basis functions transformed into the spin-orbital basis.  By a similar procedure, the derivatives of singly-excited
and doubly-excited determinants (with their corresponding CI coefficients) are
\begin{align}
        \sum_{ia} c_i^a \left| \frac{\partial \Phi_i^a}{\partial \chi} \right\rangle
    & = \sum_{ia} c_i^a \bigg[
        \left( \sum_{n \ne i} U_{nn}^{\chi} + U_{aa}^{\chi} \right) \left| \Phi_i^a \right\rangle
      + \sum_{n \ne i} \sum_{f \ne a} U_{fn}^{\chi} \left| \Phi_{in}^{af} \right\rangle \nonumber \\
    & + \sum_{f \ne a} U_{fa}^{\chi} \left| \Phi_i^f \right\rangle
      - \sum_{n \ne i} U_{in}^{\chi} \left| \Phi_n^a \right\rangle
      + U_{ia}^{\chi} \left| \Phi_0 \right\rangle \nonumber \\
    & + \sum_{n \ne i} a_{n_{\chi}}^{\dagger} a_n^{} \left| \Phi_i^a \right\rangle
      + a_{a_{\chi}}^{\dagger} a_a^{} \left| \Phi_i^a \right\rangle \bigg] \label{SingledX}
\end{align}
and
\begin{align}
        \frac{1}{4} \sum_{ijab} c_{ij}^{ab} \left| \frac{\partial \Phi_{ij}^{ab}}{\partial \chi} \right\rangle
    & = \frac{1}{4} \sum_{ijab} c_{ij}^{ab} \bigg[
        \left( \sum_{n \ne i,j} U_{nn}^{\alpha} + 2 U_{aa}^{\chi} \right) \left| \Phi_{ij}^{ab} \right\rangle
      + \sum_{n \ne i,j} \sum_{f \ne a,b} U_{fn}^{\chi} \left| \Phi_{ijn}^{abf} \right\rangle \nonumber \\
    & + 2 \sum_{f \ne a,b} U_{fa}^{\chi} \left| \Phi_{ij}^{fb} \right\rangle
      - 2 \sum_{n \ne i,j} U_{in}^{\chi} \left| \Phi_{nj}^{ab} \right\rangle
      + 4 U_{ia}^{\chi} \left| \Phi_{j}^{b} \right\rangle \nonumber \\
    & + \sum_{n \ne i,j} a_{n_{\chi}}^{\dagger} a_n^{} \left| \Phi_{ij}^{ab} \right\rangle
      + 2 a_{a_{\chi}}^{\dagger} a_a^{} \left| \Phi_{ij}^{ab} \right\rangle \bigg], \label{DoubledX}
\end{align}
respectively.  We note that when $\chi = H_\beta$, terms involving the second
quantized core derivative operators will be zero since our atomic orbital basis
functions carry no dependence on the magnetic field, i.e. we have not 
used gauge-including atomic orbitals (GIAOs) at this stage.  Inserting Eqs.~\eqref{ReferencedX}
-- \eqref{DoubledX} into Eq.~\eqref{CISDAATs1} with the relevant perturbations
and evaluating the contractions with both the standard Wick's theorem
contractions\cite{Crawford00:review}, in addition to those derived previously
\cite{Shumberger2025}, allows for an expression for the intermediately
normalized AAT to be derived.
\begin{align} \label{CISD_AAT_Intermediate_Normalization}
    [I_{\alpha\beta}^{\lambda}]_{int}
    & = \sum_m \sum_e U_{em}^{H_\beta} \left( U_{em}^{R_{\lambda\alpha}} + \left\langle \phi_m^{{R_{\lambda\alpha}}} | \phi_e^{} \right\rangle \right)
      + \sum_{ia} \frac{\partial {c_{i}^{a}}^\dagger}{\partial R_{\lambda\alpha}} U_{ai}^{H_\beta} \nonumber \\
    & - \frac{1}{2} \sum_{ia} {c_i^a}^\dagger \bigg[
        \sum_e U_{ei}^{H_\beta} \left( S_{ea}^{R_{\lambda \alpha}} + \left\langle \phi_a^{R_{\lambda \alpha}} | \phi_e \right\rangle \right)
      - \sum_m U_{am}^{H_\beta} \left( S_{im}^{R_{\lambda \alpha}} + \left\langle \phi_m^{R_{\lambda \alpha}} | \phi_i \right\rangle \right) \bigg] \nonumber \\
    & + \sum_{ia} \frac{\partial c_{i}^{a}}{\partial H_{\beta}} \left( U_{ai}^{R_{\lambda \alpha}} + \left\langle \phi_i^{R_{\lambda \alpha}} | \phi_a \right\rangle \right)
      + \sum_{ia} \frac{\partial c{_{i}^{a}}^\dagger}{\partial R_{\lambda\alpha}} \frac{\partial c_{i}^{a}}{\partial H_{\beta}} \nonumber \\
    & - \frac{1}{2} \sum_{ia} \frac{\partial c_{i}^{a}}{\partial H_{\beta}} \bigg[
        \sum_e \left( S_{ae}^{R_{\lambda \alpha}} + \left\langle \phi_e^{R_{\lambda \alpha}} | \phi_a \right\rangle \right) {c_i^e}^\dagger
      - \sum_m \left( S_{mi}^{R_{\lambda \alpha}} + \left\langle \phi_i^{R_{\lambda \alpha}} | \phi_m \right\rangle \right) {c_m^a}^\dagger \bigg] \nonumber \\
    & + \sum_{ia} c_i^a \bigg[
        U_{ia}^{H_\beta} \bigg[ \sum_{kc}
        \left( U_{kc}^{R_{\lambda \alpha}} + \left\langle \phi_c^{R_{\lambda \alpha}} | \phi_k \right\rangle \right) {c_k^c}^\dagger \bigg]
      + \sum_{me} U_{em}^{H_\beta} \bigg[
        \left( U_{em}^{R_{\lambda \alpha}} + \left\langle \phi_m^{R_{\lambda \alpha}} | \phi_e \right\rangle \right) {c_i^a}^\dagger \nonumber \\
    & - \left( U_{ei}^{R_{\lambda \alpha}} + \left\langle \phi_i^{R_{\lambda \alpha}} | \phi_e \right\rangle \right) {c_m^a}^\dagger
      - \left( U_{am}^{R_{\lambda \alpha}} + \left\langle \phi_m^{R_{\lambda \alpha}} | \phi_a \right\rangle \right) {c_i^e}^\dagger
      + \left( U_{ai}^{R_{\lambda \alpha}} + \left\langle \phi_i^{R_{\lambda \alpha}} | \phi_a \right\rangle \right) {c_m^e}^\dagger \bigg] \bigg] \nonumber \\
    & + \sum_{ijab} \frac{\partial {c_{ij}^{ab}}^\dagger}{\partial R_{\lambda\alpha}} U_{bj}^{H_\beta} c_i^a
      + \sum_{ia} \frac{\partial c_{i}^{a}}{\partial H_{\beta}} \bigg[
        \sum_{kc} \left( U_{kc}^{R_{\lambda \alpha}} + \left\langle \phi_c^{R_{\lambda \alpha}} | \phi_k \right\rangle \right) c_{ki}^{ca} \bigg] \nonumber \\
    & - \frac{1}{2} \sum_{ia} c_i^a \bigg[
      - \sum_{kme} U_{em}^{H_\beta} \left( S_{km}^{R_{\lambda \alpha}} + \left\langle \phi_m^{R_{\lambda \alpha}} | \phi_k \right\rangle \right) {c_{ki}^{ea}}^\dagger 
      + \sum_{cme} U_{em}^{H_\beta} \left( S_{ec}^{R_{\lambda \alpha}} + \left\langle \phi_c^{R_{\lambda \alpha}} | \phi_e \right\rangle \right) {c_{im}^{ac}}^\dagger \bigg] \nonumber \\
    & + \sum_{ia} \frac{\partial {c_{i}^{a}}^\dagger}{\partial R_{\lambda\alpha}} \sum_{kc} U_{kc}^{H_\beta} c_{ki}^{ca}
      + \sum_{ijab} \frac{\partial c_{ij}^{ab}}{\partial H_{\beta}} \left( U_{bj}^{R_{\lambda \alpha}} + \left\langle \phi_j^{R_{\lambda \alpha}} | \phi_b \right\rangle \right) {c_{i}^{a}}^\dagger
      + \frac{1}{4} \sum_{ijab} \frac{\partial c{_{ij}^{ab}}^\dagger}{\partial R_{\lambda\alpha}} \frac{\partial c_{ij}^{ab}}{\partial H_{\beta}} \nonumber \\
    & - \frac{1}{4} \sum_{ijab} \frac{\partial c_{ij}^{ab}}{\partial H_{\beta}} \bigg[
      - \sum_{k} \left( S_{ki}^{R_{\lambda\alpha}} + \left\langle \phi_i^{R_{\lambda\alpha}} | \phi_k^{} \right\rangle \right) {c_{kj}^{ab}}^\dagger
      + \sum_{c} \left( S_{ac}^{R_{\lambda\alpha}} + \left\langle \phi_c^{R_{\lambda\alpha}} | \phi_a^{} \right\rangle \right) {c_{ij}^{cb}}^\dagger \bigg] \nonumber \\
    & + \frac{1}{2} \sum_{ijab} {c_{ij}^{ab}}^\dagger \bigg[
        \frac{1}{2} \sum_{me} U_{em}^{H_{\beta}} \left( U_{em}^{R_{\lambda\alpha}} + \left\langle \phi_m^{R_{\lambda\alpha}} | \phi_e^{} \right\rangle \right) c_{ij}^{ab}
      - \sum_{me} U_{ej}^{H_{\beta}} \left( U_{em}^{R_{\lambda\alpha}} + \left\langle \phi_m^{R_{\lambda\alpha}} | \phi_e^{} \right\rangle \right) c_{im}^{ab} \nonumber \\
    & - \sum_{me} U_{bm}^{H_{\beta}} \left( U_{em}^{R_{\lambda\alpha}} + \left\langle \phi_m^{R_{\lambda\alpha}} | \phi_e^{} \right\rangle \right) c_{ij}^{ae} \bigg] .
\end{align}
We note that the first term in addition to the last six terms in
Eq.~\eqref{CISD_AAT_Intermediate_Normalization} would constitute the
intermediately normalized MP2/CID AAT if one chose to use first-order
amplitudes/CID coefficients and derivative first-order amplitudes/CID
coefficients in place of the CISD coefficients and derivative CISD
coefficients.  In Stephens's formulation of the AAT, it is assumed that the wave
function is fully normalized.  As such, insertion of the normalization factor
$N$ into Eq.~\eqref{CISDWaveFunction} and subsequent application to
Eq.~\eqref{ElectronicAAT} result in
\begin{align} \label{CISD_AAT_Full_Normalization}
        [I_{\alpha\beta}^{\lambda}]_{full}
    & = N^2 [I_{\alpha\beta}^{\lambda}]_{int}
      + N \frac{\partial N}{\partial R_{\lambda\alpha}} \bigg[
        2 \sum_{ia} c_i^a U_{ia}^{H_\beta}
      + \sum_{ia} c{_{i}^{a}}^\dagger \frac{\partial c_{i}^{a}}{\partial H_{\beta}} \nonumber \\
    & + 2 \sum_{ia} c_{i}^{a} \sum_{jb} U_{jb}^{H_\beta} {c_{ij}^{ab}}^\dagger
      + \frac{1}{4} \sum_{ijab} {c_{ij}^{ab}}^{\dagger} \frac{\partial c_{ij}^{ab}}{\partial H_{\beta}} \bigg]
\end{align}
for which only the last term would contribute to a fully normalized MP2/CID
formulation.  In arriving at Eqs.~\eqref{CISD_AAT_Intermediate_Normalization}
and \eqref{CISD_AAT_Full_Normalization} we have used non-canonical perturbed
orbitals which allows us to avoid numerical instabilities introduced by
degeneracies (or near-degeneracies) in the orbital energies.\cite{Handy85:MP2Hessians}  For magnetic
field perturbations, the usual relationship between the core derivative of the
overlap and CPHF coefficients, ${U_{pq}^\chi}^* + U_{qp}^\chi = S_{pq}^\chi$,
becomes ${U_{pq}^{H_\beta}}^* + U_{qp}^{H_\beta} = 0$ since the basis functions
have no dependence on magnetic fields.  Orbital rotations among the
occupied/occupied and virtual/virtual block of the CPHF coefficients result
in these blocks being zero, allowing us to remove $U_{pq}^{H_\beta}$ terms for
these subspaces from the equations. Of course, this assumes that all orbitals are active;
if we choose to freeze the core orbitals,
we must retain any terms involving $U_{pq}^{H_\beta}$ that contain core orbital indices in common with one or more CI
coefficients.  If intermediates are utilized, the terms in
Eq.~\eqref{CISD_AAT_Intermediate_Normalization} and
Eq.~\eqref{CISD_AAT_Full_Normalization} will maximally scale as $O(N^5)$. For
example, the last term in Eq.~\eqref{CISD_AAT_Intermediate_Normalization} can be
rewritten as
\begin{align}
      - \frac{1}{2} \sum_{ijab} {c_{ij}^{ab}}^\dagger \bigg[
        \sum_{me} U_{bm}^{H_{\beta}} \left( U_{em}^{R_{\lambda\alpha}} + \left\langle \phi_m^{R_{\lambda\alpha}} | \phi_e^{} \right\rangle \right) c_{ij}^{ae} \bigg]
    & = - \frac{1}{2} \sum_{ijab} {c_{ij}^{ab}}^\dagger \bigg[ \sum_e X_{be} c_{ij}^{ae} \bigg] \nonumber \\
    & = - \frac{1}{2} \sum_{ijab} {c_{ij}^{ab}}^\dagger Y_{ij}^{ab}
\end{align}
where we have reduced the scaling of the equation from $O(N^6)$ to $O(N^5)$.  We do this by first solving for
the $O(N^3)$ intermediate, $X_{be}$, which is defined as
\begin{align}
    X_{be} = \sum_{m} U_{bm}^{H_{\beta}} \left( U_{em}^{R_{\lambda\alpha}} + \left\langle \phi_m^{R_{\lambda\alpha}} | \phi_e^{} \right\rangle \right)
\end{align}
and then solving for the $O(N^5)$ intermediate $Y_{ij}^{ab}$ defined as
\begin{align}
    Y_{ij}^{ab} = \sum_e X_{be} c_{ij}^{ae}.
\end{align}
The final contraction between ${c_{ij}^{ab}}^\dagger$ and $Y_{ij}^{ab}$ is an
$O(N^4)$ step.  Solutions to the CI coefficient and their derivatives will
still scale as $O(N^6)$.

\section{Computational Details}
We have implemented the analytic-gradient method for computing CID and CISD AATs
described above in the open source Python package apyib \cite{apyib}.  This
package is supported by the Psi4 quantum chemistry package \cite{PSI4} from
which integrals and other base quantities are obtained. We compared this implementation
to a finite-difference algorithm similar to that recently
developed for MP2, \cite{Shumberger2024} but with derivatives of CI
coefficients computed separately from the determinant overlaps as expressed
in Eq.~\eqref{CISDAATs1}.  In addition, our implementation utilizes
non-canonical perturbed orbitals with the ability to invoke the frozen-core
approximation, as noted above.  

\begin{figure}[!hbtp]
    \centering
    \includegraphics[width=\textwidth]{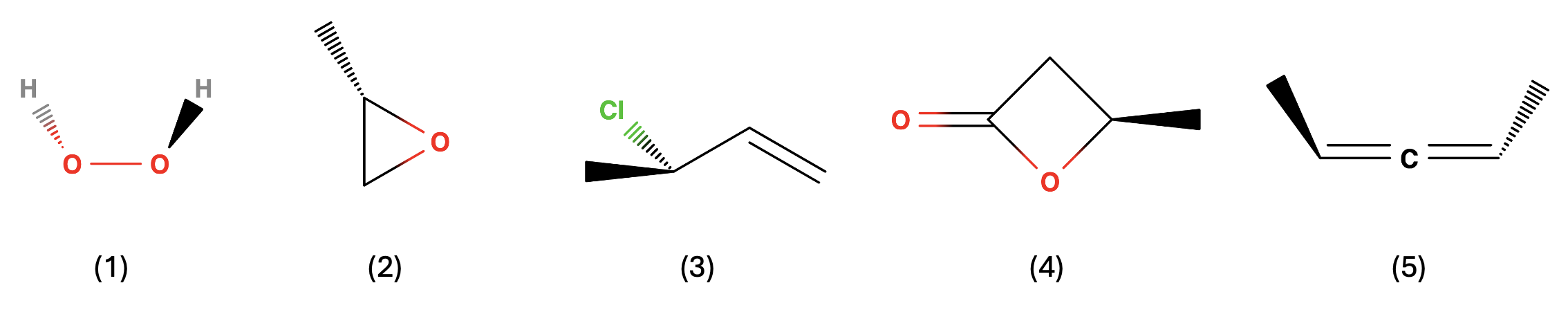}
    \caption{Molecular test set for comparisons of HF, MP2, CID, and CISD VCD spectra
    including \textbf{(1)} \hhoo, \textbf{(2)} \smo, \textbf{(3)} \rcb, 
    \textbf{(4)} \rmo, and \textbf{(5)} \dma.}
    \label{test_set}
\end{figure}

We have used this new implementation to examine the impact of the systematic
inclusion of dynamic electron correlation by drawing comparisons of VCD rotatory
strengths among the HF, MP2, CID, and CISD levels of theory computed using the
aug-cc-pVDZ basis set\cite{Dunning1989, Kendall1992a} for the molecular test set
shown in Figure \ref{test_set}.  All basis sets were obtained from the Basis Set
Exchange \cite{Pritchard2019, Feller1996, Schuchardt2007a}.  For the
finite-difference calculations the magnetic fields and geometric displacements
were set to $10^{-6}$ a.u.  Energies were converged to $10^{-13}$ a.u.\ for the
finite-difference comparisons and to $10^{-10}$ a.u.\ or greater for the
analytic evaluation of VCD quantities.  Hessians and dipole derivatives (APTs)
were obtained from the Gaussian quantum chemistry package \cite{Gaussian09}.  To
maintain a consistent comparison between VCD rotatory strengths, we used a
common-geometry/common-Hessian scheme.  Core orbitals were frozen for all
non-hydrogen atoms, and a narrow linewidth of $10^{-3}$ eV $=8.06573$
cm$^{-1}$ was used for the simulated VCD spectra.

\section{Results and Discussion}
\subsection{Comparison Between Analytic and Numerical Differentiation}

To test our implementation, we compared analytic and numerical AATs for the
hydrogen-molecule dimer, water, and \hhoo\ with various basis sets, and we found
close agreement between the two methods.  As an example, we have included the
AATs computed for \hho\ using the CID level of theory in Table \ref{h2o 6-31G}
and \hhoo\ using the CISD level of theory in Table \ref{h2o2 6-31G}, both with
the 6-31G basis set.\cite{Ditchfield1971,
Hehre1972, Hariharan1973}  Errors between the implementations were on the order of
$10^{-9}$ for \hho\ and $10^{-8}$ for \hhoo\ which show similar agreement
reported in our previous manuscript comparing the analytic approach to the
finite-difference method for the MP2 case\cite{Shumberger2025}.
\begin{table}[!hbtp]
    \scriptsize
        \caption{Electronic CID AATs (a.u.) for \hho\ computed with the 6-31G
        basis using analytic-gradient methods and finite-difference procedures.}
    \label{h2o 6-31G}
    \begin{tabular*}{\textwidth}{@{}lccccccccc}
    \toprule
    & & & \multicolumn{3}{c}{Analytic} & \hspace{1pt} & \multicolumn{3}{c}{Finite-difference} \\
    \cmidrule{4-6} \cmidrule{8-10}
    & & & $B_x$ & $B_y$ & $B_z$ & & $B_x$ & $B_y$ & $B_z$ \\
    \midrule
    O$_{1x}$ & & &  0.0000000000 & -0.1546420383 &  0.0000000000 & & -0.0000000000 & -0.1546420376 & -0.0000000008 \\
    O$_{1y}$ & & &  0.2523864912 & -0.0000000000 & -0.0000000000 & &  0.2523864918 & -0.0000000001 & -0.0000000002 \\
    O$_{1z}$ & & & -0.0000000000 & -0.0000000000 &  0.0000000000 & &  0.0000000001 & -0.0000000001 &  0.0000000006 \\
    H$_{2x}$ & & & -0.0000000000 &  0.0491462089 &  0.0714068767 & & -0.0000000000 &  0.0491462084 &  0.0714068764 \\
    H$_{2y}$ & & & -0.0208338936 &  0.0000000000 &  0.0000000000 & & -0.0208338938 &  0.0000000002 &  0.0000000001 \\
    H$_{2z}$ & & & -0.0866781888 & -0.0000000000 & -0.0000000000 & & -0.0866781886 & -0.0000000002 & -0.0000000000 \\
    H$_{3x}$ & & &  0.0000000000 &  0.0491462089 & -0.0714068767 & &  0.0000000000 &  0.0491462079 & -0.0714068775 \\
    H$_{3y}$ & & & -0.0208338936 & -0.0000000000 &  0.0000000000 & & -0.0208338942 &  0.0000000003 &  0.0000000002 \\
    H$_{3z}$ & & &  0.0866781888 &  0.0000000000 & -0.0000000000 & &  0.0866781889 & -0.0000000001 & -0.0000000001 \\
    \bottomrule
    \end{tabular*}
\end{table}

\begin{table}[!hbtp]
    \scriptsize
        \caption{Electronic CISD AATs (a.u.) for \hhoo\ computed with the 6-31G
        basis using analytic-gradient methods and finite-difference procedures.}
    \label{h2o2 6-31G}
    \begin{tabular*}{\textwidth}{@{}lccccccccc}
    \toprule
    & & & \multicolumn{3}{c}{Analytic} & \hspace{1pt} & \multicolumn{3}{c}{Finite-difference} \\
    \cmidrule{4-6} \cmidrule{8-10}
    & & & $B_x$ & $B_y$ & $B_z$ & & $B_x$ & $B_y$ & $B_z$ \\
    \midrule
    H$_{1x}$ & & & -0.1898045933 & -0.0222646607 &  0.3242355909 & & -0.1898045760 & -0.0222646630 &  0.3242355780 \\ 
    H$_{1y}$ & & & -0.0180441509 &  0.0094117839 & -0.0207843931 & & -0.0180441522 &  0.0094117811 & -0.0207843979 \\
    H$_{1z}$ & & & -0.1840522389 &  0.0929224603 &  0.1876514299 & & -0.1840522373 &  0.0929224796 &  0.1876514376 \\
    H$_{2x}$ & & & -0.1898045933 & -0.0222646607 & -0.3242355909 & & -0.1898045765 & -0.0222646557 & -0.3242355864 \\
    H$_{2y}$ & & & -0.0180441509 &  0.0094117839 &  0.0207843931 & & -0.0180441571 &  0.0094117735 &  0.0207843885 \\
    H$_{2z}$ & & &  0.1840522389 & -0.0929224603 &  0.1876514299 & &  0.1840522317 & -0.0929224552 &  0.1876514226 \\
    O$_{3x}$ & & & -0.0451458652 &  0.0794663609 &  1.1847772282 & & -0.0451458859 &  0.0794663660 &  1.1847772322 \\
    O$_{3y}$ & & & -0.1508788119 & -0.0042901597 &  0.0160726896 & & -0.1508787987 & -0.0042901616 &  0.0160726887 \\
    O$_{3z}$ & & & -1.1567011160 & -0.0219982861 &  0.0422548261 & & -1.1567011357 & -0.0219982849 &  0.0422548271 \\
    O$_{4x}$ & & & -0.0451458652 &  0.0794663609 & -1.1847772282 & & -0.0451458794 &  0.0794663859 & -1.1847772241 \\
    O$_{4y}$ & & & -0.1508788119 & -0.0042901597 & -0.0160726896 & & -0.1508788143 & -0.0042901346 & -0.0160726874 \\
    O$_{4z}$ & & &  1.1567011160 &  0.0219982861 &  0.0422548261 & &  1.1567011203 &  0.0219982969 &  0.0422548247 \\ 
    \bottomrule
    \end{tabular*}
\end{table}

\subsection{VCD Spectrum of \hhoo}
\begin{figure}[!hbtp]
    \centering
    \includegraphics[width=\textwidth]{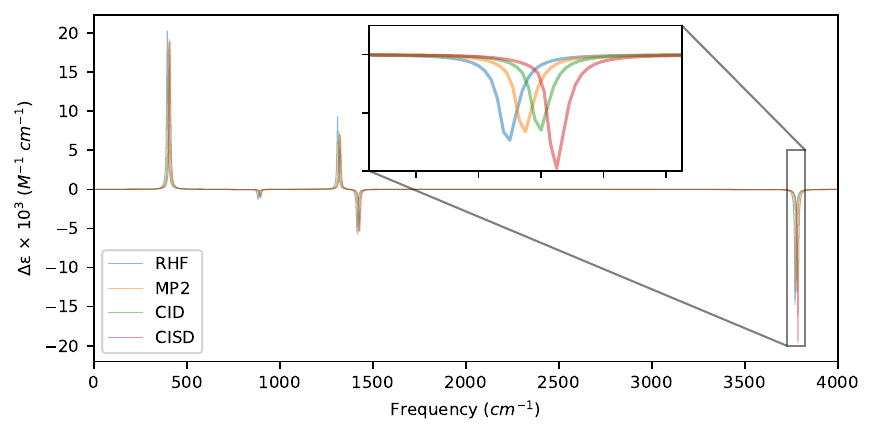}
    \caption{VCD spectra of \hhoo\ computed at the HF, MP2, and CISD levels of theory with
             aug-cc-pVDZ using a common optimized geometry and Hessian obtained at the
             MP2/aug-cc-pVDZ level of theory. Artificial shifts of 5 cm$^{-1}$, 10 cm$^{-1}$,
             and 15 cm$^{-1}$ for MP2, CID, and CISD, respectively, have been introduced into the spectrum
             to distinguish the peaks more easily.}
    \label{h2o2_aug-cc-pVDZ_vcd}
\end{figure}

Using the recently developed analytic derivative formulation for computing MP2
AATs and the CID/CISD AAT approaches derived here, we compare VCD spectra and the
requisite rotatory strengths for \hhoo\ using the HF, MP2, CID, and CISD levels of
theory.  We find rather distinct differences between the four spectra even for
a small molecule such as \hhoo.  The vibrational frequencies and corresponding
rotatory strengths are presented in Table \ref{h2o2_aug-cc-pVDZ_rotstr} and
the VCD spectrum in Figure \ref{h2o2_aug-cc-pVDZ_vcd}.  
\begin{table}[!hbtp]
    \small
    \caption{VCD rotatory strengths of \hhoo\ computed at the HF, MP2, CID, and CISD
             levels of theory with aug-cc-pVDZ using a common optimized geometry and
             Hessian obtained at the MP2/aug-cc-pVDZ level.}
    \sisetup{table-format = 2.4, table-alignment-mode = format}
    \label{h2o2_aug-cc-pVDZ_rotstr}
    \begin{tabular*}{\textwidth}{@{\extracolsep{\fill}}
        S[table-alignment-mode = none]
            c   
        S[table-number-alignment = right]
        S[table-number-alignment = right]
        S[table-number-alignment = right]
        S[table-number-alignment = right]
        @{}}
    \toprule
{Frequency} & \hspace{1pt} & \multicolumn{4}{c}{{Rotatory Strength}} \\
{(cm$^{-1}$)} & & \multicolumn{4}{c}{{($10^{-44}$ esu$^2$ cm$^2$)}} \\
    \cmidrule{0-0} \cmidrule{3-6}
    & & {HF} & {MP2} & {CID} & {CISD} \\
    \midrule
     3768.80 & & -17.909 & -17.970 & -17.416 & -29.793 \\ 
     3767.46 & &   7.091 &   8.738 &   8.312 &  17.149 \\
     1414.00 & & -11.330 & -12.111 & -10.977 & -10.995 \\
     1309.92 & &  20.802 &  15.368 &  15.984 &  15.344 \\
      881.29 & &  -4.304 &  -2.759 &  -3.454 &  -3.688 \\
      394.22 & & 150.336 & 138.916 & 142.114 & 139.937 \\
    \bottomrule
    \end{tabular*}
\end{table}

The most notable differences between the HF and MP2 levels of theory occur at
394.22 cm$^{-1}$ and 1309.92 cm$^{-1}$ which are related to hydrogen bending
motions perpendicular and parallel to the O$-$O bond axis, respectively.  For
these vibrational modes, the MP2 rotatory strengths are predicted to be $\sim
11 \times 10^{-44}$ esu$^2$ cm$^2$ and $\sim 5 \times 10^{-44}$ esu$^2$ cm$^2$
smaller than those from the uncorrelated wave function.  CID predicts similar
rotational strengths, however, optimization of the CI coefficients shifts the
rotational strengths for the 394.22 cm$^{-1}$ mode back in the direction of
that produced by the HF wave function.  Likewise, the CISD rotatory strengths
for these modes are similar to those predicted by the MP2 method.
Interestingly, we note substantial differences in rotatory strengths between
the MP2 and CID methods versus CISD for the nearly degenerate in-phase and
out-of-phase hydrogen bond stretches at 3767.46 cm$^{-1}$ and 3768.80
cm$^{-1}$, respectively.  Both of these modes exhibit rotatory strengths of
almost an order of magnitude more intense for the CISD method over that of MP2
and CID --- a direct consequence of including singly excited determinants.
The differences between these nearly degenerate modes are not as noticeable in
the spectra due to the fact that the peaks have opposite signs leading to some
cancellation between them, depending on the spectral resolution, and as shown
in the inset of Figure \ref{h2o2_aug-cc-pVDZ_vcd}.

\subsection{VCD Spectrum of \smo}

\begin{figure}[!hbtp]
    \centering
    \includegraphics[width=\textwidth]{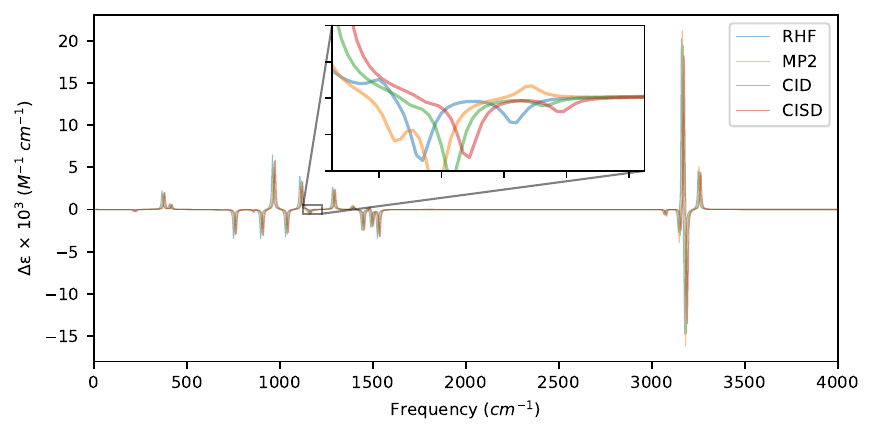}
    \caption{VCD spectra of \smo\ computed at the HF, MP2, and CISD levels of theory with
             aug-cc-pVDZ using a common optimized geometry and Hessian obtained at the 
             MP2/aug-cc-pVDZ level of theory. Artificial shifts of 5 cm$^{-1}$, 10 cm$^{-1}$,
             and 15 cm$^{-1}$ for MP2, CID, and CISD, respectively, have been introduced into the spectrum
             to distinguish the peaks more easily.}
    \label{smo_aug-cc-pVDZ_vcd}
\end{figure}

In Table \ref{smo_aug-cc-pVDZ_rotstr} and Figure \ref{smo_aug-cc-pVDZ_vcd}, we
present the rotatory strengths and VCD spectra of \smo\ computed at the HF,
MP2, CID, and CISD levels of theory. Two vibrational modes are noteworthy:
1183.14 cm$^{-1}$, which is a rocking motion of all the hydrogen atoms, and
1140.36 cm$^{-1}$ the mode associated with a bending motion of the hydrogen
atoms on the CH$_2$ carbon atom parallel to the plane of the epoxide ring.
The former exhibits a sign discrepancy between HF, CID, and CISD versus MP2,
while the latter exhibits a similar discrepancy between HF and the correlated
methods, as shown in the inset of Figure \ref{smo_aug-cc-pVDZ_vcd}.  Both
modes produce rotatory strengths less than $1 \times 10^{-44}$ esu$^2$ cm$^2$
for each of the methods tested which given their relative weakness makes the
sign discrepancies relatively unsurprising.  Interestingly, the modes most
affected by the inclusion of singly excited determinants into the wave
function are hydrogen stretching motions 3158.52 cm$^{-1}$ and 3175.86
cm$^{-1}$.  In these modes, we note a shift in rotatory strength of about
$1-1.5 \times 10^{-44}$ esu$^2$ cm$^2$ between the CID and CISD methods
providing evidence for a small contribution to the rotatory strengths and VCD
spectra from inclusion of singly excited determinants into the wave function
for \smo.

\begin{table}[!hbtp]
    \small
    \caption{VCD rotatory strengths of \smo\ computed at the HF, MP2, CID, and CISD 
             levels of theory with aug-cc-pVDZ using a common optimized geometry and
             Hessian obtained at the MP2/aug-cc-pVDZ level.}
    \sisetup{table-format = 2.4, table-alignment-mode = format}
    \label{smo_aug-cc-pVDZ_rotstr}
    \begin{tabular*}{\textwidth}{@{\extracolsep{\fill}}
        S[table-alignment-mode = none]
            c
        S[table-number-alignment = right]
        S[table-number-alignment = right]
        S[table-number-alignment = right]
        S[table-number-alignment = right]
        @{}}
    \toprule
{Frequency} & \hspace{1pt} & \multicolumn{4}{c}{{Rotatory Strength}} \\
{(cm$^{-1}$)} & & \multicolumn{4}{c}{{($10^{-44}$ esu$^2$ cm$^2$)}} \\
    \cmidrule{0-0} \cmidrule{3-6}
    & & {HF} & {MP2} & {CID} & {CISD} \\
    \midrule
     3248.50 & &   4.202 &   4.640 &   4.090 &   4.010 \\ 
     3175.86 & & -14.496 & -15.784 & -14.379 & -13.225 \\
     3158.52 & &  23.028 &  24.315 &  22.291 &  20.602 \\
     3154.34 & &  -5.751 &  -6.226 &  -5.666 &  -4.926 \\
     3141.40 & &  -3.124 &  -4.287 &  -3.421 &  -2.923 \\
     3063.53 & &  -0.566 &  -0.674 &  -0.624 &  -0.765 \\
     1523.43 & &  -6.659 &  -5.626 &  -6.184 &  -6.236 \\
     1488.22 & &  -4.004 &  -4.054 &  -3.893 &  -3.642 \\
     1471.44 & &   0.770 &   0.880 &   0.798 &   0.755 \\
     1439.22 & &  -5.082 &  -5.262 &  -5.109 &  -5.057 \\
     1384.01 & &   0.750 &   1.153 &   0.967 &   0.810 \\
     1283.08 & &   6.265 &   5.715 &   5.746 &   5.583 \\
     1183.14 & &  -0.453 &   0.226 &  -0.137 &  -0.250 \\
     1153.29 & &  -1.208 &  -1.688 &  -1.472 &  -1.113 \\
     1140.36 & &   0.279 &  -0.724 &  -0.130 &  -0.103 \\
     1106.88 & &  10.859 &   7.497 &   9.396 &   9.032 \\
     1029.37 & &  -9.447 &  -8.055 &  -8.583 &  -7.938 \\
      959.19 & &  20.484 &  15.893 &  18.032 &  18.510 \\
      896.28 & & -11.327 &  -8.763 &  -9.988 & -10.073 \\
      843.85 & &  -0.107 &  -0.815 &  -0.528 &  -1.170 \\
      750.52 & & -13.478 & -10.745 & -11.862 & -11.393 \\
      405.53 & &   4.797 &   4.325 &   4.445 &   4.187 \\
      365.86 & &  17.634 &  14.357 &  15.986 &  16.124 \\
      212.36 & &  -3.652 &  -3.346 &  -3.427 &  -3.497 \\
    \bottomrule
    \end{tabular*}
\end{table}

\subsection{VCD Spectrum of \rcb}

\begin{figure}[!hbtp]
    \centering
    \includegraphics[width=\textwidth]{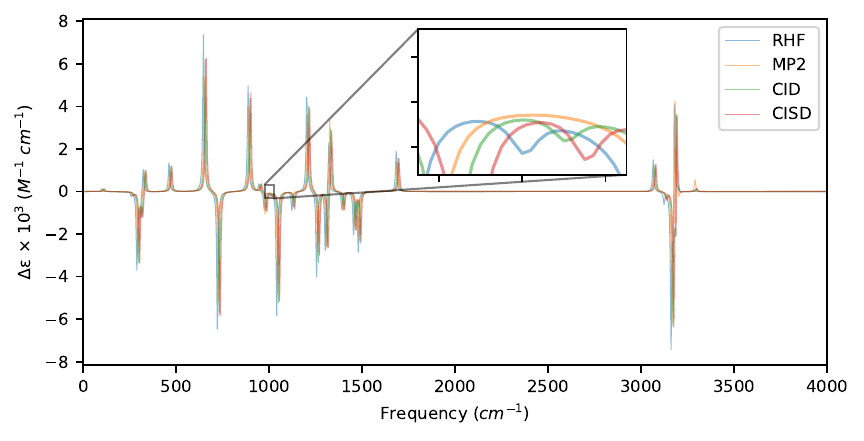}
    \caption{VCD spectra of \rcb\ computed at the HF, MP2, and CISD levels of theory with
             aug-cc-pVDZ using a common optimized geometry and Hessian obtained at the 
             MP2/aug-cc-pVDZ level of theory. Artificial shifts of 5 cm$^{-1}$, 10 cm$^{-1}$,
             and 15 cm$^{-1}$ for MP2, CID, and CISD, respectively, have been introduced into the spectrum
             to distinguish the peaks more easily.}
    \label{rcb_aug-cc-pVDZ_vcd}
\end{figure}

The rotatory strengths and VCD spectra of \rcb\ are reported in Table
\ref{rcb_aug-cc-pVDZ_rotstr} and Figure \ref{rcb_aug-cc-pVDZ_vcd},
respectively.  Of the thirty vibrational modes associated with \rcb, only two
modes exhibit sign changes among the four levels of theory examined here.  In
the first mode at 1000.49 cm$^{-1}$, a hydrogen bending vibration, we note a
sign change of the MP2 method relative to all others, as illustrated by the
inset of Figure \ref{rcb_aug-cc-pVDZ_vcd}.  Likewise, the vibrational mode at
3195.37 cm$^{-1}$, a C$-$H stretching vibration of hydrogens associated with
the C$=$C double bond, deviates in its sign at the HF level relative to CID,
MP2, and CISD.  However, both of these vibrations exhibit very weak rotatory
strengths, similarly to the case of \smo\ above.  Additionally, two modes at
1254.14 cm$^{-1}$ and 885.18 cm$^{-1}$, the former a hydrogen bending motion
on the Cl$-$C$-$H bond angle and the latter a more complex mode involving
combinations of C$-$H bending and C$-$C stretching motions, are the the most
affected by the inclusion of singly excited determinants into the wave
function, though the shift is small --- less than $1 \times 10^{-44}$ esu$^2$
cm$^2$ relative to CID method, indicating that singly excited determinants do
not significantly contribute to the rotatory strengths of \rcb.

\begin{table}[!hbtp]
    \small
    \caption{VCD rotatory strengths of \rcb\ computed at the HF, MP2, CID, and CISD
             levels of theory with aug-cc-pVDZ using a common optimized geometry and
             Hessian obtained at the MP2/aug-cc-pVDZ level.}
    \sisetup{table-format = 2.4, table-alignment-mode = format}
    \label{rcb_aug-cc-pVDZ_rotstr}
    \begin{tabular*}{\textwidth}{@{\extracolsep{\fill}}
        S[table-alignment-mode = none]
            c   
        S[table-number-alignment = right]
        S[table-number-alignment = right]
        S[table-number-alignment = right]
        S[table-number-alignment = right]
        @{}}
    \toprule
{Frequency} & \hspace{1pt} & \multicolumn{4}{c}{{Rotatory Strength}} \\
{(cm$^{-1}$)} & & \multicolumn{4}{c}{{($10^{-44}$ esu$^2$ cm$^2$)}} \\
    \cmidrule{0-0} \cmidrule{3-6}
    & & {HF} & {MP2} & {CID} & {CISD} \\
    \midrule
     3283.33 & &   0.064 &   0.514 &   0.127 &   0.132 \\ 
     3195.37 & &   0.194 &  -0.292 &  -0.044 &  -0.092 \\
     3175.65 & &  13.563 &  11.129 &  10.946 &  10.663 \\
     3173.97 & & -12.168 &  -8.545 &  -9.140 &  -8.890 \\
     3159.73 & &  -6.780 &  -5.733 &  -5.817 &  -5.880 \\
     3122.02 & &  -0.507 &  -0.127 &  -0.273 &  -0.341 \\
     3063.13 & &   1.486 &   1.204 &   1.284 &   1.266 \\
     1681.99 & &   3.274 &   2.372 &   2.716 &   2.695 \\
     1481.56 & &   1.397 &   0.497 &   0.892 &   0.671 \\
     1477.73 & &  -6.342 &  -4.651 &  -5.263 &  -4.956 \\
     1452.78 & &  -4.101 &  -3.285 &  -3.624 &  -3.674 \\
     1389.58 & &  -1.786 &  -1.776 &  -1.850 &  -1.790 \\
     1320.00 & &   5.363 &   7.708 &   6.757 &   6.552 \\
     1300.95 & &  -6.692 &  -6.321 &  -6.356 &  -6.494 \\
     1254.14 & &  -9.338 &  -6.557 &  -7.771 &  -6.986 \\
     1201.01 & &  11.477 &   9.319 &  10.305 &  10.100 \\
     1120.95 & &  -2.376 &  -1.171 &  -1.968 &  -2.198 \\
     1040.08 & & -16.312 & -13.395 & -14.644 & -14.381 \\
     1000.49 & &  -0.484 &   0.005 &  -0.331 &  -0.578 \\
      971.74 & &  -2.537 &  -3.236 &  -2.705 &  -2.864 \\
      943.63 & &   1.132 &   1.091 &   1.048 &   1.154 \\
      885.18 & &  17.036 &  13.795 &  15.014 &  15.902 \\
      720.96 & & -27.735 & -21.947 & -24.329 & -24.977 \\
      645.30 & &  34.303 &  24.976 &  28.860 &  29.071 \\
      460.50 & &   8.695 &   6.600 &   7.447 &   7.795 \\
      321.73 & &  10.527 &   9.282 &   9.946 &   9.977 \\
      304.83 & & -11.310 &  -9.569 & -10.611 & -10.526 \\
      287.72 & & -37.090 & -31.644 & -33.486 & -33.682 \\
      257.56 & &  -2.016 &  -1.216 &  -1.678 &  -1.718 \\
       97.78 & &   3.750 &   3.321 &   3.729 &   3.709 \\
    \bottomrule
    \end{tabular*}
\end{table}

\subsection{VCD Spectrum of \rmo}

\begin{figure}[!hbtp]
    \centering
    \includegraphics[width=\textwidth]{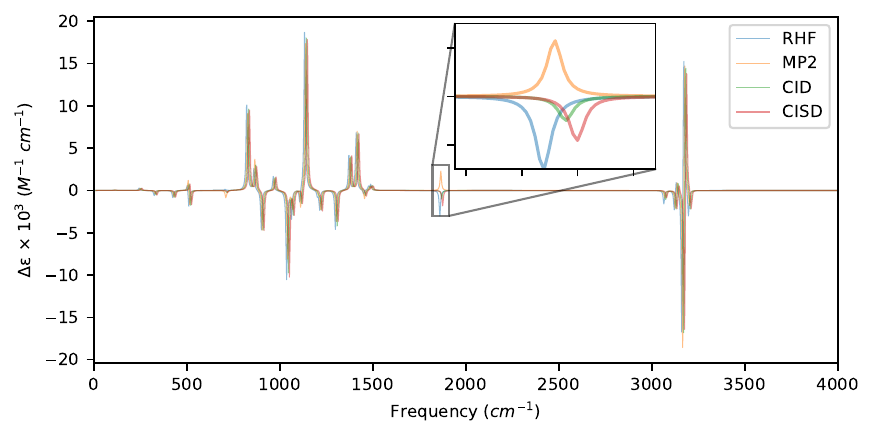}
    \caption{VCD spectra of \rmo\ computed at the HF, MP2, and CISD levels of theory with
             aug-cc-pVDZ using a common optimized geometry and Hessian obtained at the 
             MP2/aug-cc-pVDZ level of theory. Artificial shifts of 5 cm$^{-1}$, 10 cm$^{-1}$,
             and 15 cm$^{-1}$ for MP2, CID, and CISD, respectively, have been introduced into the spectrum
             to distinguish the peaks more easily.}
    \label{rmo_aug-cc-pVDZ_vcd}
\end{figure}

Table \ref{rmo_aug-cc-pVDZ_rotstr} and Figure \ref{rmo_aug-cc-pVDZ_vcd}
present the rotatory strengths and VCD spectra for \rmo, respectively.  We
find two interesting peaks with sign discrepancies between the methods.  The
most significant sign discrepancy of our test set, a carbonyl stretch at
1859.68 cm$^{-1}$, is predicted by the MP2 method to have a rotational
strength of $+3.611 \times 10^{-44}$ esu$^2$ cm$^2$ while HF, CID, and CISD
predicted negative values of $-4.749 \times 10^{-44}$ esu$^2$ cm$^2$, $-1.560
\times 10^{-44}$ esu$^2$ cm$^2$, and $-2.856 \times 10^{-44}$ esu$^2$ cm$^2$,
respectively, as shown in the inset of Figure \ref{rmo_aug-cc-pVDZ_vcd}.
Likewise, the inclusion of singly excited determinants shifts the rotational
strength from by about $-1.3 \times 10^{-44}$ esu$^2$ cm$^2$ from the CID to
CISD methods --- one of the more significant deviations observed thus far.
The second mode, a ring breathing motion at 704.03 cm$^{-1}$, is predicted by
the HF method to have a positive sign while the MP2, CID, and CISD methods
predict negative signs, though this mode is much weaker.  In addition to the
mode at 1859.68 cm$^{-1}$, we observe two other modes affected by the
inclusion of singly excited determinants at 1297.72 cm$^{-1}$ and 1035.89
cm$^{-1}$, the former a C$-$H bending motion and the latter a bending motion
involving both the C$=$O double bond and the methyl group. The shifts in the
rotatory strengths between the CID and CISD methods are between $1 - 1.5
\times 10^{-44}$ esu$^2$ for both modes, indicating, again, only a slight
effect of the inclusion of singly excited determinants into the wave function
for \rmo.

\begin{table}[!hbtp]
    \small
    \caption{VCD rotatory strengths of \rmo\ computed at the HF, MP2, CID, and CISD
             levels of theory with aug-cc-pVDZ using a common optimized geometry and
             Hessian obtained at the MP2/aug-cc-pVDZ level.}
    \sisetup{table-format = 2.4, table-alignment-mode = format}
    \label{rmo_aug-cc-pVDZ_rotstr}
    \begin{tabular*}{\textwidth}{@{\extracolsep{\fill}}
        S[table-alignment-mode = none]
            c
        S[table-number-alignment = right]
        S[table-number-alignment = right]
        S[table-number-alignment = right]
        S[table-number-alignment = right]
        @{}}
    \toprule
{Frequency} & \hspace{1pt} & \multicolumn{4}{c}{{Rotatory Strength}} \\
{(cm$^{-1}$)} & & \multicolumn{4}{c}{{($10^{-44}$ esu$^2$ cm$^2$)}} \\
    \cmidrule{0-0} \cmidrule{3-6}
    & & {HF} & {MP2} & {CID} & {CISD} \\
    \midrule
     3195.99 & &  -2.934 &  -1.788 &  -2.173 &  -2.146 \\ 
     3171.17 & &  16.723 &  16.350 &  15.917 &  15.277 \\
     3160.57 & & -17.647 & -19.338 & -17.636 & -17.197 \\
     3130.64 & &   1.347 &   1.327 &   1.228 &   0.960 \\
     3119.72 & &  -2.239 &  -2.057 &  -2.129 &  -2.073 \\
     3063.98 & &  -1.495 &  -0.834 &  -0.994 &  -0.806 \\
     1859.68 & &  -4.749 &   3.611 &  -1.560 &  -2.856 \\
     1487.08 & &   1.422 &   0.761 &   1.099 &   0.910 \\
     1476.00 & &   0.853 &   0.760 &   0.877 &   0.946 \\
     1449.43 & &  -0.990 &  -2.184 &  -1.410 &  -1.575 \\
     1409.39 & &  14.335 &  14.727 &  14.300 &  14.046 \\
     1370.99 & &   9.175 &   8.432 &   8.800 &   9.055 \\
     1297.72 & & -10.422 &  -8.312 &  -9.500 &  -8.333 \\
     1213.11 & &  -5.836 &  -5.691 &  -5.739 &  -6.115 \\
     1201.03 & &  -1.503 &  -0.950 &  -1.263 &  -0.824 \\
     1132.78 & &  50.039 &  46.530 &  48.258 &  47.860 \\
     1106.25 & &  -4.509 &  -6.141 &  -4.948 &  -5.026 \\
     1061.42 & &  -8.951 &  -5.813 &  -7.699 &  -7.669 \\
     1035.89 & & -29.483 & -22.713 & -27.232 & -28.667 \\
      964.31 & &   5.144 &   4.310 &   4.761 &   4.960 \\
      899.66 & & -15.386 & -15.445 & -15.453 & -15.613 \\
      860.03 & &   8.614 &  12.292 &   9.961 &   9.435 \\
      821.12 & &  37.429 &  33.642 &  35.619 &  35.198 \\
      704.03 & &   0.565 &  -3.580 &  -1.029 &  -1.022 \\
      509.75 & & -11.831 & -11.399 & -11.395 & -10.904 \\
      501.83 & &   5.994 &   9.434 &   6.999 &   6.756 \\
      424.19 & &  -5.861 &  -5.242 &  -5.458 &  -5.474 \\
      323.72 & &  -5.418 &  -3.786 &  -4.710 &  -4.903 \\
      242.97 & &   3.460 &   3.056 &   3.240 &   3.277 \\
      103.73 & &   2.033 &   2.081 &   2.028 &   2.018 \\
    \bottomrule
    \end{tabular*}
\end{table}

\subsection{VCD Spectrum of \dma}

\begin{figure}[!hbtp]
    \centering
    \includegraphics[width=\textwidth]{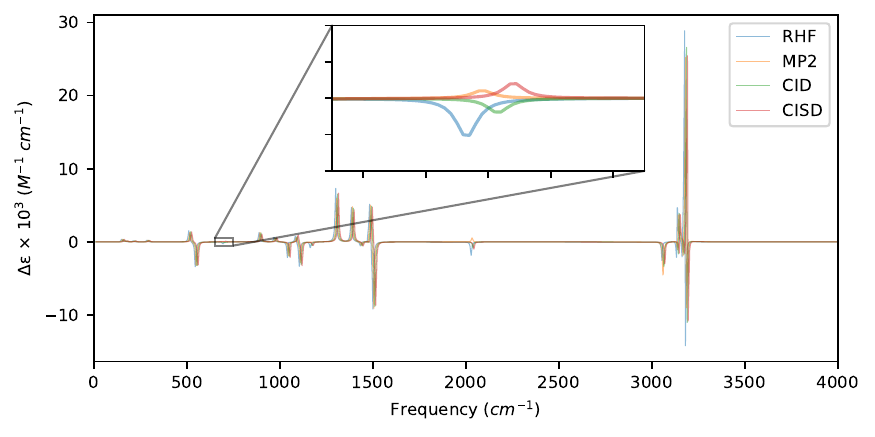}
    \caption{VCD spectra of \dma\ computed at the HF, MP2, and CISD levels of theory with
             aug-cc-pVDZ using a common optimized geometry and Hessian obtained at the
             MP2/aug-cc-pVDZ level of theory. Artificial shifts of 5 cm$^{-1}$, 10 cm$^{-1}$,
             and 15 cm$^{-1}$ for MP2, CID, and CISD, respectively, have been introduced into the spectrum
             to distinguish the peaks more easily.}
    \label{dma_aug-cc-pVDZ_vcd}
\end{figure}

In Table \ref{dma_aug-cc-pVDZ_rotstr} and Figure \ref{dma_aug-cc-pVDZ_vcd}, we
present the rotatory strengths and VCD spectra for \dma.  In contrast to the
other molecules in our test set, the inclusion of singly excited determinants
plays a significant role in the spectroscopic simulation of VCD for \dma.
Most notable is the vibrational mode at 693.06 cm$^{-1}$ --- a mode associated
with a bending motion of the hydrogen atoms on the allene moiety.
Interestingly, HF and CID methods predict this mode to exhibit a negative
rotatory strength while the MP2 and CISD levels of theory predicted a positive
value, though all four methods predict small values, as illustrated by the
inset of Figure \ref{dma_aug-cc-pVDZ_vcd}.

The intense pairs of modes at 3134.69/3134.81 cm$^{-1}$ and at 3176.99/3177.99
cm$^{-1}$ exhibit the largest absolute shifts in rotatory strengths between
CID and CISD, as well as the largest shifts due to dynamic electron
correlation effects.  These rotatory strengths for these modes, which
correspond to in-phase/out-of-phase C$-$H stretches (the former pair on the
cumulene chain and the latter on the methyl groups), occur in opposite-sign
pairs and thus exhibit significant cancellation in the VCD spectra.  The
inclusion of single excitations between CID and CISD reduces the magnitude of
the rotatory strengths of the methyl C$-$H stretches by $3-4 \times 10^{-44}$
esu$^2$ cm$^{2}$ and those of the cumulenic C$-$H stretches by $6-7 \times
10^{-44}$ esu$^2$ cm$^{2}$.  Additionally, the shift in the rotatory strengths
between HF and the correlated methods is even larger: for the methyl C$-$H
stretches the intensities are reduced in magnitude by ca.\ $28 \times
10^{-44}$ esu$^2$ cm$^{2}$, while that for the cumulenic C$-$H stretched by
almost $59 \times 10^{-44}$ esu$^2$ cm$^{2}$ between HF and MP2.  We observe
an additional sign discrepancy for the weak mode at 2027.70 cm$^{-1}$ which
corresponds to vibration of the two C$=$C double bonds.  For this mode, HF,
CID, and CISD predict the mode to have a negative sign between $-2.672 \times
10^{-44}$ esu$^2$ and $-1.192 \times 10^{-44}$ esu$^2$ while MP2 predicts a
positive value of $0.813 \times 10^{-44}$ esu$^2$.

\begin{table}[!hbtp]
    \small
    \caption{VCD rotatory strengths of \dma\ computed at the HF, MP2, CID, and CISD
             levels of theory with aug-cc-pVDZ using a common optimized geometry and 
             Hessian obtained at the MP2/aug-cc-pVDZ level.}
    \sisetup{table-format = 2.4, table-alignment-mode = format}
    \label{dma_aug-cc-pVDZ_rotstr}
    \begin{tabular*}{\textwidth}{@{\extracolsep{\fill}}
        S[table-alignment-mode = none]
            c   
        S[table-number-alignment = right]
        S[table-number-alignment = right]
        S[table-number-alignment = right]
        S[table-number-alignment = right]
        @{}}
    \toprule
{Frequency} & \hspace{1pt} & \multicolumn{4}{c}{{Rotatory Strength}} \\
{(cm$^{-1}$)} & & \multicolumn{4}{c}{{($10^{-44}$ esu$^2$ cm$^2$)}} \\
    \cmidrule{0-0} \cmidrule{3-6}
    & & {HF} & {MP2} & {CID} & {CISD} \\
    \midrule
     3177.99 & & -123.004 & -95.171 & -105.805 & -102.292 \\ 
     3176.99 & &  133.216 & 105.986 &  116.353 &  112.238 \\
     3163.83 & &    6.123 &   1.049 &    3.132 &    2.963 \\
     3163.03 & &   -9.216 &  -4.356 &   -6.139 &   -5.800 \\
     3134.81 & &  184.936 & 126.346 &  150.405 &  143.942 \\
     3134.69 & & -183.138 &-124.610 & -148.863 & -142.436 \\
     3056.10 & &    8.233 &   7.031 &    7.259 &    6.440 \\
     3055.54 & &  -10.203 & -11.287 &  -10.229 &   -9.082 \\
     2027.70 & &   -2.672 &   0.813 &   -1.192 &   -1.435 \\
     1499.71 & &  -18.434 & -18.252 &  -17.764 &  -17.398 \\
     1481.80 & &   10.876 &  10.617 &   10.310 &   10.058 \\
     1471.33 & &    6.098 &   6.054 &    5.855 &    5.355 \\
     1470.28 & &   -5.718 &  -5.742 &   -5.498 &   -4.987 \\
     1431.97 & &   -1.201 &  -1.019 &   -1.175 &   -1.214 \\
     1385.40 & &   -7.362 &  -9.321 &   -7.999 &   -7.456 \\
     1384.16 & &   16.421 &  18.446 &   16.876 &   15.914 \\
     1299.51 & &   16.623 &  13.499 &   14.832 &   15.114 \\
     1163.04 & &   -2.035 &  -0.137 &   -1.237 &   -1.273 \\
     1103.82 & &   -8.886 &  -7.539 &   -8.344 &   -8.670 \\
     1083.04 & &    2.535 &   1.951 &    2.367 &    2.933 \\
     1043.14 & &    4.624 &   3.466 &    3.939 &    4.202 \\
     1040.68 & &   -9.120 &  -7.653 &   -8.233 &   -8.799 \\
      968.74 & &    1.970 &   1.551 &    1.716 &    1.774 \\
      887.61 & &    4.281 &   4.279 &    4.067 &    3.582 \\
      824.20 & &    0.195 &   0.096 &    0.127 &    0.145 \\
      693.05 & &   -1.124 &   0.231 &   -0.416 &    0.446 \\
      544.89 & &  -19.123 & -18.133 &  -18.168 &  -17.939 \\
      509.78 & &    9.067 &   7.686 &    8.171 &    8.120 \\
      284.38 & &    2.411 &   2.322 &    2.316 &    2.083 \\
      210.94 & &    1.915 &   1.617 &    2.099 &    1.853 \\
      167.53 & &    2.366 &   1.940 &    2.069 &    2.113 \\
      148.40 & &    7.089 &   6.985 &    6.617 &    6.764 \\
      129.93 & &   -1.014 &  -1.015 &   -0.958 &   -1.025 \\
    \bottomrule
    \end{tabular*}
\end{table}

\section{Conclusions}

Based on a second-quantized formalism for derivatives of Slater determinants,
we have obtained expressions for the first time for CID- and CISD-level AATs.
Our formulation uses non-canonical perturbed orbitals, allows for frozen-core
orbitals, and scales as ${\cal O}(N^5)$.  We have validated the
analytic-derivative implementation against corresponding numerical derivatives
and observed excellent agreement.  Additionally, we have compared HF, MP2,
CID, and CISD rotatory strengths and VCD spectra for \hhoo, \smo, \rcb, \rmo,
and \dma\ using the aug-cc-pVDZ basis.  For four of the five test molecules,
we observed sign differences between the HF and MP2 methods relative to CID
and CISD.  Similarly, the inclusion of singly excited determinants into the
wave function for \hhoo\ and \dma\ significantly affects the rotatory
strengths of peaks in the high-frequency region, though near degeneracy of the
corresponding modes suppresses these effects somewhat in the resulting
spectra.  Additionally, we observed a sign change between CID and CISD for
\dma\ for a weak mode in the mid-IR region.

Curiously, we observed that the molecules most affected by the inclusion of
singly excited determinants into the wave function are those which exhibit
axial chirality (\hhoo\ and \dma) rather than point chirality (\smo, \rcb, and
\rmo), though determining whether this is a general phenomenon will require
additional investigation.  We will consider this topic in future work, as well
as the important issue of gauge-origin dependence of the rotatory strengths.


\clearpage

\section{Supporting Information} \label{si}

Atomic coordinates of the test molecules are provided.

\section{Acknowledgements} \label{ack}

TDC was supported by the U.S.\ National Science Foundation via grant CHE-2154753 and BMS by grant DMR-1933525.
The authors are grateful to Advanced Research Computing at Virginia Tech for providing computational resources
that contributed to the results reported within the paper.

\bibliography{references}

\newpage

\end{document}